\documentclass[a4paper,12pt]{article}
\pdfoutput=1

\usepackage{mystyle} 

\usepackage[T1]{fontenc} 
\usepackage{tikz,caption,subcaption} 
\usetikzlibrary{decorations.markings}

\usepackage{epsfig,euscript,xspace, color}
\usepackage{mathtools}

\def\bea{\begin{eqnarray}}
\def\eea{\end{eqnarray}}
\def\be{\begin{equation}}
\def\ee{\end{equation}}

\usetikzlibrary{arrows,snakes}

\definecolor{lightblue}{rgb}{.1,.4,.5}
\definecolor{brown1}{rgb}{.64,.43,.138}

\usepackage{mathrsfs}  

\title{\boldmath Light cone OPE in a CFT with lowest twist scalar primary}

\author{Atanu Bhatta, Soham Ray} 
\affiliation{Indian Institute of Science\\ Bengaluru  560012, India 
}

\emailAdd{atanubhatta@iisc.ac.in}
\emailAdd{sohamray@iisc.ac.in}

\abstract{We study the operator product expansion (OPE) of two identical scalar primary operators in the light cone limit in a conformal field theory where a scalar is the operator with lowest twist. We see that in CFTs where both the stress tensor and a scalar are the lowest twist operators, the stress tensor  contributes at the leading order in the light cone OPE and  the scalar contributes at the subleading order. We also see that there does not exist a scalar analog of the average null energy condition for a CFT where a scalar is the lowest twist operator.}

\begin{document} 

\maketitle
\flushbottom

\section{INTRODUCTION}
The operator product expansion (OPE) is used often in nonperturbative studies of quantum field theories. It states that the product of two nearby operators can be replaced by a sum over all the operators present in the theory. The OPE has all the information of the corresponding theory. In a conformal field theory (CFT) the OPE takes the following form
\begin{equation}
\mathcal{O}_1(x)\mathcal{O}_2(0) = \sum_{\mathcal{O}} \left. C_{\mathcal{O}_1\mathcal{O}_2\mathcal{O}} (x,\partial_y) \mathcal{O}(y) \right|_{y=0} 
\end{equation}
where $C_{\mathcal{O}_1\mathcal{O}_2\mathcal{O}} (x,\partial_y)$ is a power series in $\partial_y$ and the sum is over all primary operators $\mathcal{O}$. In recent time it has become very useful to study the OPE in the light cone limit. Given two operators in a CFT, the light cone limit is taken in a way such that the two operators approach the light cone and become null separated. The OPE of two such operators is known as a light cone OPE~\cite{Brandt:1970kg}. In a CFT, the light cone OPE is dominated by the operators of the lowest twist, where twist of an operator is $\tau = \Delta - l$, $\Delta$ and $l$ being the mass dimension and spin of the operator, respectively~\cite{Komargodski:2012ek}. The light cone OPE can be inserted inside a CFT correlator to compute the conformal blocks which can be used in the bootstrap programs~\cite{Fitzpatrick:2014vua, Simmons-Duffin:2016wlq,Fitzpatrick:2012yx,Beem:2015aoa}. The universality of the lowest twist operator light cone OPE is studied in~\cite{Fitzpatrick:2019zqz,Huang:2019fog, Fitzpatrick:2019efk}.

 In a seminal work,  Hartman {\it et al.} found that the light cone OPE of two identical scalar primary operators in a CFT with the stress tensor $T_{\mu\nu}$ as the lowest twist operator has a global contribution from the so-called average null energy operator defined as
 \begin{equation}
 \mathcal{E} = \int_{-\infty}^{\infty} du'~ T_{uu}(u',v=0)
 \end{equation}
 where $u, v$ are the light cone coordinates~\cite{Hartman:2016lgu}.
Inserting the light cone OPE into correlation functions and studying their analyticity properties it can be proved that the averaged null energy condition (ANEC) - one of the important statements of general relativity - holds in a CFT~\cite{Borde:1987qr,Roman:1988vv,Hartman:2015lfa,Hartman:2016lgu}. The ANEC is written as the following operator statement in CFTs:
\begin{equation}
	\langle \Psi|\mathcal{E}|\Psi\rangle \geq 0
\end{equation}
This means that the averaged null energy operator $\mathcal{E}$ has a non-negative expectation value in any arbitrary state $\Psi$. This condition is also proved in the theory of quantum information using the monotonicity of relative entropy~\cite{Faulkner:2016mzt}. There exist some other proofs of the ANEC as well which hold only in free or superrenormalizable theories or in two dimensions~\cite{Klinkhammer:1991ki,Wald:1991xn,Folacci:1992xg,Verch:1999nt,Bousso:2015wca}. The ANEC in the CFTs with higher spin lowest twist operators is also discussed in~\cite{Hartman:2016lgu}.

However there are CFTs with the scalar primary operators having the lowest twist, for example, the long-range Ising model, the $\phi^{3}$ theory in $6-\epsilon$ dimensions and the $\mathcal{N}=4$ supersymmetric Yang-Mills (SYM) theory in four dimensions. In the long-range Ising model~\cite{Paulos:2015jfa,Rychkov:2016mrc,Behan:2017dwr,Behan:2017emf,Behan:2018hfx} and the $\phi^{3}$ theory in $6-\epsilon$ dimensions the lowest twist operator is a scalar~\cite{Gopakumar:2016cpb,Goncalves:2018nlv}. For  $\mathcal{N}=4$ SYM, both the stress tensor and a scalar primary operator that belong to the short multiplet have the lowest twist~\cite{Beem:2013sza,Beem:2013qxa,Beem:2016wfs}. In this paper, we study the light cone OPE in such CFTs and try to see if one can obtain a scalar analog of the ANEC. We also see that a scalar analog of the ANEC does not exist in a CFT where the scalar is the lowest twist operator. 

This paper is organized as follows. In Sec. ~\ref{sec:review} we describe the light cone limit and the light cone OPE. We also show how the light cone OPE can be used to prove the ANEC. In Sec. ~\ref{sec:ope} we study the light cone OPE in a CFT where a scalar is the lowest twist operator. We pick a specific example of a theory where a scalar is the lowest twist operator - the long-range Ising model. We also study the $\mathcal{N}=4$ supersymmetric Yang-Mills theory in $d=4$ which has both the stress tensor and a scalar primary as lowest twist operators. We show that at leading order only the stress tensor contributes to the light cone OPE and not the scalar.  In Sec. ~\ref{sec:phi3} we show that we cannot have a condition analogous to the ANEC for scalars. We finally list our findings and conclusions in Sec. ~\ref{sec:conclusions}.
%
%
%
%
%
\section{\label{sec:review}GENERALITIES}
In this section, we briefly describe the light cone OPE in a CFT and its use to obtain ANEC from causality. We shall work in the light cone coordinates $(u,v,{\bf x})$ defined as
\begin{equation}
u=t-y\,;\quad v=t+y
\end{equation}
where $(t,y,{\bf x})$ are the Cartesian coordinates which describe the Minkowski spacetime $\mathbb{R}^{1,d-1}$.
Let us consider two identical scalar primary operators $\psi(u,v,{\bf 0})$ and $\psi(-u,-v,{\bf 0})$ in a CFT. Since we restrict ourselves in the ${\bf x}= {\bf 0}$ plane in the rest of the paper, from now on we denote $\psi(u,v,{\bf 0})$ as $\psi(u,v)$ unless otherwise mentioned. In the light cone limit, these two operator insertions approach the light cone. 
\vskip 0.5 cm
\begin{center}
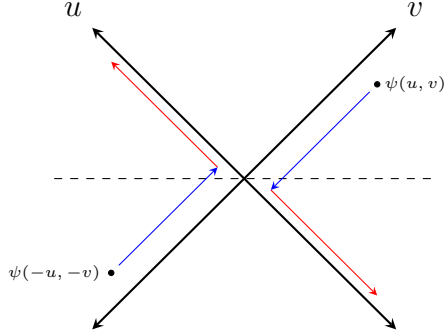

	\begin{tikzpicture} [scale=0.5]
	 \draw [dashed] (-5,0) -- (5,0);
	 \draw [>= stealth,thick, ->] (0,0) -- (4,4);
	 \draw [>= stealth,thick, ->] (0,0) -- (-4,-4);
	 \draw [>= stealth,thick, ->] (0,0) -- (-4,4);
	 \draw [>= stealth,thick, ->] (0,0) -- (4,-4);
	 \filldraw [black] (3.5,2.5) circle (2pt);
	 \filldraw [black] (-3.5,-2.5) circle (2pt);
	 \draw [>= stealth,blue, ->] (3.3,2.3) -- (0.7,-0.3);
	 \draw [>= stealth,blue, ->] (-3.3,-2.3) -- (-0.7,0.3);
	 \draw [>= stealth,red, ->]  (0.7,-0.3) -- (3.5,-3.1);
	 \draw [>= stealth,red, ->]  (-0.7,0.3) -- (-3.5,3.1);
	 \node at (4.5,4.5) {$v$};
	 \node at (-4.5,4.5) {$u$};
	 \node at (4.5,2.5) {\tiny $  \psi(u,v)$};
	 \node at (-5,-2.5) {\tiny $  \psi(-u,-v)$};
	\end{tikzpicture}
	\captionof{figure}{Light cone limit}
	\label{fig1}
\end{center}
For example, as shown in Fig.~\ref{fig1} we first take the limit $v \rightarrow 0$ and then the limit $u \rightarrow \infty$. Thus, in the light cone limit, we have:
\begin{equation}
 |v| \ll \frac{1}{|u|} \ll 1
\end{equation}
In this limit, the distance between $\psi(u,v)$ and $\psi(-u,-v)$ is $\sqrt{4 uv} \ll 1$, i.e., they are null separated.

The OPE of two primary operators in the light cone limit is dominated by the operators of the lowest twist~\cite{Komargodski:2012ek}. If a CFT has the stress tensor to be the lowest twist operator, then in the $v\to 0$ limit, the leading order contribution in the OPE comes from the following components of the stress tensor and its descendants, 
$T_{uu},\partial_{u}T_{uu},\partial_{u}^{2}T_{uu},...$. This allows us to write the light cone OPE in the following summation form
\begin{equation}
\frac{\psi(u,v)\psi(-u,-v)}{\langle\psi(u,v)\psi(-u,-v)\rangle} = 1+vu^{3}\sum_{n=0}^{\infty}c_{n}(u \partial_{u})^{n} T_{uu}(0)
\end{equation}
where $c_{n}$ are constants to be determined from the dynamics of the CFT. It is more useful to see the light cone OPE in the following integral form~\cite{Hartman:2016lgu}
\begin{equation}
\label{eq:ope-integral}
\frac{\psi(u,v)\psi(-u,-v)}{\langle\psi(u,v)\psi(-u,-v)\rangle} = 1 - \frac{15 c_{\psi\psi T}}{c_{T}} vu^{2} \int_{-u}^{u}du^{'} \left(1-\frac{u^{'2}}{u^{2}}\right)^{2} T_{uu}(u^{'},v=0)
\end{equation}
where the constants $c_{\psi\psi T}$ and $c_T$ are the dynamical quantities that appear in the correlators $\langle \psi(u,v)\psi(-u,-v) T_{uu}(u_{3},v_{3}) \rangle$ and $\langle T_{uu}(u,0) T_{uu}(u_{3},v_{3}) \rangle$~\cite{Osborn:1993cr}. Notice the kernel in~\eqref{eq:ope-integral}, $ \left(1-\frac{u^{'2}}{u^{2}}\right)^{2}$ has been chosen such that at large $u$ it goes to $1$ and it is consistent with the three-point functions.
Finally we take the $u\to \infty$ limit as mentioned before to get the light cone OPE as
\begin{equation}
\label{eq:nullenergyinope}
\frac{\psi(u,v)\psi(-u,-v)}{\langle\psi(u,v)\psi(-u,-v)\rangle} = 1 + \lambda_{T} vu^{2} \mathcal{E}
\end{equation}
where $\lambda_{T}$ is a positive constant and
\begin{equation}
 \mathcal{E} = \int_{-\infty}^{\infty} du'~ T_{uu}(u',0)
\end{equation}
is the average null-energy operator.
\begin{center}
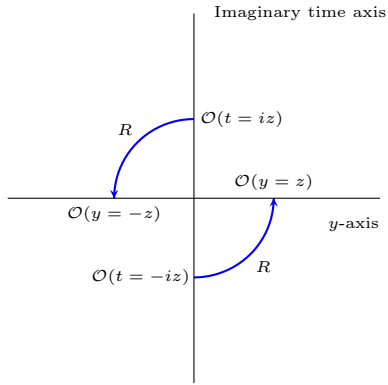

	\begin{tikzpicture}[scale=0.7]
	\draw [blue,thick,domain=270:360] plot ({1.5*cos(\x)}, {1.5*sin(\x)});
	\draw [blue,thick,domain=90:180] plot ({1.5*cos(\x)}, {1.5*sin(\x)});
	\draw [>= stealth,blue, ->] (1.5,-0.001)--(1.5,0);
	\draw [>= stealth,blue, ->] (-1.5,0.001)--(-1.5,0);
	\draw (0,3.5) -- (0,-3.5);
	\draw (-3.5,0) -- (3.5,0);
	\node at (2, 3.5) { \tiny Imaginary time axis};
	\node at (3,-0.5) {\tiny $y$-axis};
	\node at (1.3,-1.3) {\tiny $R$};
	\node at (-1.3,1.3) {\tiny $R$};
	\node at (0.9,1.5) {\tiny $\mathcal{O}(t=iz)$};
	\node at (-1.5,-0.3) {\tiny $\mathcal{O}(y=-z)$};
	\node at (1.5,0.3) {\tiny $\mathcal{O}(y=z)$};
	\node at (-1,-1.5) {\tiny $\mathcal{O}(t=-iz)$};
	\end{tikzpicture}
	\captionof{figure}{Action of rotation operator $R$.}
	\label{fig2}
\end{center}
One can use the light cone OPE to get an ANEC in the CFT.  In~\cite{Hartman:2016lgu,Hartman:2016dxc,Hartman:2015lfa}, it is shown that for an arbitrary operator $O$ (not necessarily primary) the ANEC becomes 
\begin{equation}
\label{eq:anec}
\langle\mathcal{O}^{\dagger}(t=i\delta)\mathcal{E}\mathcal{O}(t=-i\delta)\rangle \geq 0
\end{equation}
where $\mathcal{O} = R.O$ with the rotation operator $R$ which causes a rotation by $\frac{\pi}{2}$ in the $\tau-y$ plane where $\tau = i t$, as shown in Fig. ~\ref{fig2}. The proof is outlined in Appendix~\ref{app:anecproof}.
%
%
%
%
%
%
\section{\label{sec:ope}LIGHT CONE OPE WITH THE LOWEST TWIST SCALAR PRIMARY}
Now we look for a light cone OPE when a scalar is the lowest twist operator in a CFT. The twist of a scalar primary $\phi$ is $\tau = \Delta_{\phi}$ where $\Delta_{\phi}$ is the conformal weight of the scalar primary. For a scalar, it is evident that its descendants also have the same twist. As we are taking $v\to 0$ limit, naturally, the $v$ indices of the descendants will not contribute to the OPE at leading order. Thus, the operators that appear in the light cone OPE are $\phi,\partial_{u}\phi,\partial_{u}^{2}\phi,...$ and so on. Thus, in this case, the summation form of the light cone OPE becomes 
\begin{equation}
\label{eq:scalaropesum}
\frac{\psi(u,v)\psi(-u,-v)}{\langle \psi(u,v)\psi(-u,-v) \rangle } = 1 + (uv)^{\Delta_{\phi}/2}\sum_{n=0}^{\infty} c_{n} (u\partial_{u})^{n}\phi(u,0)
\end{equation}
To find an integral from of the light cone OPE in this case, we begin with the following ansatz 
\begin{equation}
\label{eq:scalaropeansatz}
\frac{\psi(u,v)\psi(-u,-v)}{\langle \psi(u,v)\psi(-u,-v) \rangle} = 1 + f(u,v) \int_{-u}^{u} du' K(u,u') \, \phi(u',0)
\end{equation}
We demand the kernel $K(u,u')$ should go to $1$ in the limit $u \rightarrow \infty$ as in \eqref{eq:ope-integral}. Looking at the term  $(uv)^{\tau/2}$ in \eqref{eq:scalaropesum}, we choose $f(u,v) = \frac{(uv)^{\Delta_{\phi}/2}}{u}$, and write \eqref{eq:scalaropeansatz} as
\begin{equation}
\label{eq:light coneopeint}
\frac{\psi(u,v)\psi(-u,-v)}{\langle \psi(u,v)\psi(-u,-v) \rangle } = 1 + \lambda\frac{(uv)^{\Delta_{\phi}/2}}{u} \int_{-u}^{u} du' K(u,u') \phi(u',0)
\end{equation}
where $\lambda$ is a dimensionless dynamical constant that can be determined from the relevant correlators. To obtain the kernel, we have to consider following CFT correlation functions:
\begin{eqnarray}
\langle\phi(u,0)\phi(u_{3},v_{3})\rangle &=& \frac{1}{[(u-u_{3})(-v_{3})]^{\Delta_{\phi}}} \\
\label{eq:scalarthreept}
\langle\psi(u,v)\psi(-u,-v)\phi(u_{3},v_{3})\rangle &=& \frac{c_{\psi\psi\phi}}{(4uv)^{\Delta_{\psi}-\frac{\Delta_{\phi}}{2}}[(u^{2}-u_{3}^{2})(-v_{3}^{2})]^{\frac{\Delta_{\phi}}{2}}} \\
\langle\psi(u,v)\psi(-u,-v)\rangle &=& \frac{1}{(4uv)^{\Delta_{\psi}}}
\end{eqnarray}
These correlators are exact solutions of the conformal Ward identities, i.e., they are fixed by the conformal symmetries up to some multiplicative constants.
Now, we use \eqref{eq:light coneopeint} to compute the three-point function $\langle\psi(u,v)\psi(-u,-v)\phi(u_{3},v_{3})\rangle$ and comparing with the right-hand side of~\eqref{eq:scalarthreept}, we determine the kernel to be
\begin{equation}
\label{eq:scalarkernel}
K(u,u') = \left( 1-\frac{u'^2}{u^2}\right)^{\Delta_{\phi}/2 -1}
\end{equation}
The kernel for lowest twist operators with general spin is available in~\cite{Hartman:2016lgu}.  The dynamical constant $\lambda$ introduced in~\eqref{eq:light coneopeint} is proportional to the OPE coefficient $c_{\psi\psi\phi}$ and from now on we shall denote $\lambda$ as $\lambda_{\psi\psi\phi}$.
Finally, we take the limit $u \to \infty$ to write the light cone OPE as
\begin{eqnarray}
\frac{\psi(u,v)\psi(-u,-v)}{\langle\psi(u,v)\psi(-u,-v)\rangle} &=& 1 + \lambda_{\psi\psi\phi} \frac{(uv)^{\Delta_{\phi}/2}}{u} \int_{-\infty}^{\infty} du'  \left( 1-\frac{u'^2}{u^2}\right)^{\Delta_{\phi}/2 -1} \phi(u',0) \cr 
&\equiv& 1 + \lambda_{\psi\psi\phi} \frac{(uv)^{\Delta_{\phi}/2}}{u}  \mathcal{S}
\label{eq:finallight coneope}
\end{eqnarray}
Recall that in the light cone limit $uv$ is very small as compared to $1$ and $u \rightarrow \infty$. This helps us to see that when the correction term from the scalars in the light cone OPE is in subleading order or negligible in comparison to the identity term whenever the mass dimension $\Delta_{\phi}$ is positive, i.e., the CFT is unitary. Notice that the expression for the light cone OPE~\eqref{eq:finallight coneope} is valid for any scalar primary $\psi$. Let us now consider two specific examples of conformal field theories in which scalar primaries are the lowest twist operators.
\subsection{Long-range Ising model in $d=3$}
The long-range Ising (LRI) model is a unitary theory where there is no conserved stress tensor at the fixed point~\cite{Paulos:2015jfa,Rychkov:2016mrc,Behan:2017dwr,Behan:2017emf,Behan:2018hfx}. In $d$ dimension, the theory is described by the Hamiltonian,
\begin{equation}
H_{LRI} = -J \sum_{ij} \frac{S_i S_j}{|i-j|^{d+s}}
\end{equation}
where $1\leq d < 4$~\cite{Paulos:2015jfa}. We take $s$ to be $d/2 <s <s_*$ where at $s = s_*$ the theory becomes short-range Ising model. In this domain of $s$ the theory describes a nontrivial universal fixed point whose field theoretic description is given by
\begin{equation}
S= \int d^dx \, \frac{1}{2}\phi (-\partial^2)^{s/2} \phi + \frac{g}{4!} \,\phi^4(x)
\end{equation}
where the scalar field $\phi$ represents the spin density and has the conformal weight $\Delta_{\phi} = (d-s)/2$. All other composite operators can be built from $\phi$. However, one can have a rank two symmetric tensor as
\begin{equation}
 T_{\mu\nu} = \left( \phi \partial_\mu \partial_\nu \phi -\frac{1}{d} \eta_{\mu\nu} \phi \partial^2 \phi \right) - \frac{\Delta_{\phi}+1}{\Delta_{\phi}} \left( \partial_\mu\phi \partial_\nu\phi -\frac{1}{d} \eta_{\mu\nu} (\partial \phi)^2\right)
\end{equation}
This is not conserved when $s < s_*$. At $s=s_*$, this becomes the conserved stress tensor of the (short-range) Ising model. The operator dimension $\Delta_T$ of the leading spin-2 operator $T_{\mu\nu}$ has been computed in~\cite{Behan:2018hfx}, and it was shown that $\Delta_T$ approaches $d$ as $s \to s_*$. In $d=3$, for   $3/2 <s <s_*$, $\Delta_T$ varies between $3$ and $3.5$~\cite{Behan:2018hfx}.

In $d=3$, since $s> 3/2$, the twist of the scalar operator,
\begin{equation}
 \tau_\phi = \Delta_{\phi} = \frac{3-s}{2} < \frac{3}{4}
\end{equation} 
The twist of $T_{\mu\nu}$ in $d=3$ varies between $1$ and $1.5$. Therefore for $3/2 <s <s_*$, the scalar operator $\phi$ is the lowest twist operator.

However the action possesses $\mathbb{Z}_2$ global symmetry. The operator $\phi$ is $\mathbb{Z}_2$ odd. The OPE $\psi\psi$ we are considering is always $\mathbb{Z}_2$ even. Therefore, $\phi$ does not appear in the OPE. The $\mathbb{Z}_2$ even operators with lowest twist are $\phi^2$ and $T_{\mu\nu}$. Near $s \to d/2$, they both have twist, $\tau \approx 3/2$. At the leading order, both the operators contribute to the light cone OPE. The contribution from $O=\phi^2$ in the light cone OPE of two identical scalars becomes
\begin{eqnarray}
\frac{\psi(u,v)\psi(-u,-v)}{\langle\psi(u,v)\psi(-u,-v)\rangle} &=& 1 + \lambda_{\psi\psi O} \frac{(vu)^{(3-s)/2}}{u} \int_{-\infty}^{\infty} du' \, O(u',0) \cr 
&\equiv& 1+ \lambda_{\psi\psi O} \frac{(vu)^{(3-s)/2}}{u} \mathcal{S} 
\end{eqnarray}
Note that at $u\to\infty$ the contribution from $\mathcal{S}$ becomes negligible. 
\subsection{$\mathcal{N} =4$ supersymmetric Yang-Mills theory}
Let us consider a CFT in which there are two operators of lowest twist - the stress tensor and a scalar. For example, the $\mathcal{N} =4$ supersymmetric Yang-Mills theory in $d=4$. The stress tensor $T_{\mu\nu}$ and the scalar primary $\mathcal{O}_2$, both belong to the short multiplet and have the lowest twist $\tau =2$. For this case, the light cone OPE takes the following form
\begin{eqnarray}
\frac{\psi(u,v)\psi(-u,-v)}{\langle \psi(u,v)\psi(-u,-v) \rangle } = 1 &+& \lambda_{\psi\psi T} vu^{2} \int_{-u}^{u} du_{1} K_{1}(u,u_{1})  T_{uu}(u_{1},0) \cr 
&+&  \lambda_{\psi\psi \mathcal{O}}v \int_{-u}^{u} du_{2} K_{2}(u,u_{2})  \mathcal{O}_2(u_{2},0)
\label{eq:opesym}
\end{eqnarray}
As before, the kernels $K_{1}$ and $K_{2}$ are to be determined by considering the three-point correlators $\langle \psi\psi\mathcal{O}_2 \rangle$ and $ \langle \psi\psi T_{uu} \rangle$. However, note that the two-point function $\langle\mathcal{O}_2 T_{uu}\rangle$ vanishes. Thus, when we consider the correlator $ \langle \psi\psi T_{uu} \rangle$, the third term in the right-hand side of~\eqref{eq:opesym} drops out and we find the kernel $K_1$ to match with that given in~\eqref{eq:ope-integral}. On taking the correlator with $\langle \psi\psi\mathcal{O}_2 \rangle$ we found that there is no contribution from  the second term in the right hand side of~\eqref{eq:opesym}. From the contribution of the third term, we obtain the kernel $K_2$ and see that it matches with ~\eqref{eq:scalarkernel}. Therefore, it demonstrates if a CFT has both the stress tensor and a scalar primary having the lowest twist, only the stress tensor contributes in the light cone OPE at the leading order, whereas the scalar do at the subleading orders.
%
%
%
%
%
\section{\label{sec:phi3}SCALAR ANALOG OF ANEC}
Although we have seen in the previous section that the contribution from $\mathcal{S}$ in the light cone OPE is negligible in a unitary CFT, one can still search for an ANEC-like condition for $\mathcal{S}$.\footnote{However for nonunitary CFTs, Rindler positivity  does not hold~\cite{Casini:2010bf,Maldacena:2015waa}. }If the stress tensor or any spin-2 operator share the lowest twist along with scalar primaries, one will have an ANEC for the spin-2 operators. For a unitary theory, with a scalar primary as the only lowest twist operator, we now check if there is a condition for $\mathcal{S}$  like the ANEC for $\mathcal{E}$.  
  
One might expect that the condition for $\mathcal{S}$ will look like
\begin{equation}
\label{eq:suff_cond}
i \lambda_{\psi\psi\phi} \langle\overline{O(y=\delta)} \mathcal{S} O(y=\delta)\rangle \geq 0
\end{equation}
just replacing $\lambda_{T}$ by $\lambda_{\psi\psi\phi}$ and $\mathcal{E}$ by $\mathcal{S}$ in\eqref{eq:anec}. However, to investigate if this is true we start with the general form of the light cone OPE,
\begin{equation}
\label{eq:genlight coneope}
\frac{\psi(u,v)\psi(-u,-v)}{\langle\psi(u,v)\psi(-u,-v)\rangle} = 1 + \lambda_{\psi\psi\phi} u^n v^m \mathcal{S}
\end{equation}
From~\eqref{eq:finallight coneope}, we have $m=\Delta_{\phi}/2$ and $n = \Delta_{\phi}/2-1$ when a scalar primary is the only lowest twist operator in a CFT. Now, we consider the following correlator:
\begin{equation}
	G = \frac{\langle\overline{O(y=\delta)}\psi(u,v)\psi(-u,-v)O(y=\delta)\rangle}{\langle\overline{O(y=\delta)}O(y=\delta)\rangle\langle\psi(u,v)\psi(-u,-v)\rangle}.
\end{equation}
We make the following change of variables:
\begin{equation}
v = -\eta \sigma \quad u = 1/\sigma
\end{equation}
In the $\sigma$ plane, $G(\sigma)$ is analytic in the region enclosed by the contour shown in the Fig. ~\ref{fig3} and $Re(G) \leq |G| \leq 1$ (for further details, see Appendix~\ref{app:anecproof}).
 \begin{center}
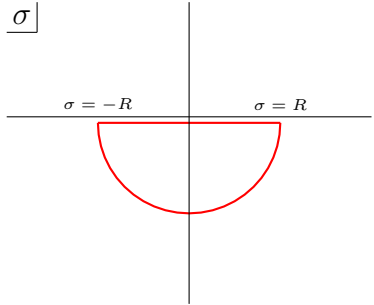

 	\begin{tikzpicture}[scale=0.8]
 	\draw [red,thick,domain=180:360] plot ({1.5*cos(\x)}, {1.5*sin(\x)});
 	\draw [red,thick] (-1.5,0) -- (1.5,0);
 	\draw (-3,0.1) -- (3,0.1);
 	\draw (0,2) -- (0,-3);
 	\draw (-2.5,1.5) -- (-2.5,2);
 	\draw (-3,1.5) -- (-2.5,1.5);
 	\node at (-1.5,0.3) {\tiny $\sigma=-R$};
 	\node at (1.5,0.3) {\tiny $\sigma=R$};
 	\node at (-2.75,1.75) { $\sigma$};
 	\end{tikzpicture}
 	\captionof{figure}{Integration contour for $G(\sigma)$. The correlation function is analytic in the above disc region. The semicircle is the closed contour over which we integrate.}
 	\label{fig3}
 \end{center}
Therefore
\begin{equation}
	\oint d\sigma (1-G(\sigma)) = 0
\end{equation}
Using the light cone OPE~\eqref{eq:genlight coneope}, we find
\begin{equation}
1-G(\sigma) = (-1)^{m+1} \frac{\lambda_{\psi\psi\phi}}{N_\delta} \eta^m \sigma^{m-n} \langle \overline{O(y=\delta)}\mathcal{S}O(y=\delta)\rangle
\end{equation}
where $N_{\delta} = \langle \overline{O(y=\delta)}O(y=\delta) \rangle$. Performing the contour integration, we get 
\begin{equation}
\label{eq:contint}
\langle \overline{O(y=\delta)}\mathcal{S}O(y=\delta)\rangle \frac{e^{-i\pi(m-n+1)}-1}{i(m-n+1)} =\frac{N_{\delta}}{\lambda_{\psi\psi\phi} \eta^m} \int_{-R}^{R}d\sigma Re(1-G(\sigma))
\end{equation}
up to an overall sign. $G(\sigma)$ has the property that $Re(G) \le 1$ (see Appendix \ref{app:anecproof}). This makes the right-hand side of~\eqref{eq:contint} non-negative,
\begin{equation}
\label{eq:nnrhs}
 \frac{N_{\delta}}{\lambda_{\psi\psi\phi} \eta^m} \int_{-R}^{R}d\sigma Re(1-G(\sigma)) \ge 0
\end{equation}
 For a bound like the ANEC to be possible (i.e., for $\mathcal{S}$ to have a definite sign), we must have $m-n+1 = 0$.
But, we have
\begin{eqnarray}
m-n+1 &=& (\Delta_\phi/2) -(\Delta_\phi/2 -1)  + 1 =2 \neq 0
\end{eqnarray}
This implies 
\begin{equation}
 \frac{e^{-i\pi(m-n+1)}-1}{i(m-n+1)} =0
\end{equation}
This means the right-hand side of~\eqref{eq:contint} is forced to be zero saturating the bound in~\eqref{eq:nnrhs}. Thus, we cannot impose any condition like~\eqref{eq:suff_cond} on the expectation value $\langle \overline{O(y=\delta)}\mathcal{S}O(y=\delta)\rangle$ appeared in the left-hand side of~\eqref{eq:contint}.
Therefore, we see that there is no ANEC-like condition in any CFT where a scalar is the unique lowest twist operator.
%
%
%
%
\section{\label{sec:conclusions}SUMMARY AND CONCLUSIONS}
We have studied the light cone OPE in a unitary CFT where the lowest twist operator is a scalar primary operator. There exist some CFTs where  both the stress tensor (or any rank two symmetric tensor primary) and a scalar primary have lowest twist, for example the $\mathcal{N}=4$ supersymmetric Yang-Mills theory in $d=4$.  We show that in such cases the stress tensor contributes to the light cone OPE at leading order whereas the scalar primary contributes at the subleading order.

Using the above findings, we conclude that for a unitary CFT where a scalar is the lowest twist operator, we cannot have a scalar analog of the ANEC. However the ANEC involving the stress tensor still exists in such theory even if the contribution from the stress tensor appears at the subleading order in the light cone OPE. 

One cannot use the Rindler positivity condition for a nonunitary theory. Thus, our method to obtain ANEC fails in such theories. It will be interesting to propose a methodology to find ANEC-like bounds in the nonunitary CFTs, for example, the IR fixed point of $\phi^3$ theory in $d=6-\epsilon$ dimensions. 

It will be interesting to study the light cone OPE for the CFTs in which the lowest twist operator is neither the stress tensor nor a scalar operator, say, for instance, a conserved current, and to check for the existence of analog of ANEC-like bounds.


\acknowledgments
{}
We thank Aninda Sinha for guiding and helping us at several steps in our work. We are also grateful to Sandipan Kundu for useful correspondence.

\appendix
%
\section{PROOF OF ANEC}
\label{app:anecproof}
We now summarize the proof of the ANEC as given in ~\cite{Hartman:2016lgu}. Recall the light cone OPE ~\eqref{eq:nullenergyinope}.
Consider the correlator,
\begin{equation}
	G_{anec} = \langle\overline{O}\psi(u,v)\psi(-u,-v)O\rangle
\end{equation}
The bar here denotes Rindler reflection~\cite{Casini:2010bf,Maldacena:2015waa}, which is reflection across the origin of the $u-v$ plane. Notice that $\psi(u,v)$ is thus the Rindler reflection of $\psi(-u,-v)$, which we denote as $\psi$.
Using the inner product $(A,B) = \langle\overline{A}B\rangle$ and using the Cauchy Schwartz inequality, we have
\begin{equation}
	Re (G_{anec}) \leq |G_{anec}| \leq (\langle\overline{O \psi} O \psi\rangle\langle\overline{\psi O} \psi O\rangle)^{1/2}
\end{equation}
As both the terms inside the square root are in Rindler positive ordering (see~\cite{Hartman:2015lfa,Hartman:2016lgu}) it is dominated by the Euclidean OPE in which we can shift the terms around. This allows us to write
\begin{equation}
	\langle\overline{O \psi} O \psi\rangle \sim \langle\overline{\psi O} \psi O\rangle \sim \langle\overline{O} O\rangle\langle\overline{\psi} \psi\rangle
\end{equation}
Thus, we have
\begin{equation}
Re (G_{anec}) \leq |G_{anec}| \leq \langle\overline{O} O\rangle\langle\psi(u,v) \psi(-u,-v)\rangle + \epsilon
\end{equation}
$\epsilon$ here is a correction term which can be neglected in the light cone limit.
We now change coordinates to
\begin{equation}
	v = -\eta \sigma \quad u = 1/\sigma
\end{equation}
Now, consider
\begin{equation}
\label{eq:anec_correlator}
	G = \frac{\langle\overline{O(y=\delta)}\psi(u,v)\psi(-u,-v)O(y=\delta)\rangle}{\langle\overline{O(y=\delta)}O(y=\delta)\rangle\langle\psi(u,v)\psi(-u,-v)\rangle}
\end{equation}
$G(\sigma)$ follows two important properties:
\begin{enumerate}
	\item For normalized operators $O$ and $\sigma$, $Re(G) \leq |G| \leq 1$.
	
	\item $G$ is analytic in a small region near the origin in the lower half plane of $\sigma$. This follows from the fact that the correlator $\langle O_{1}(x_{1})...O_{n}(x_{n}) \rangle$ is analytic when we have $Im(x_{1})\triangleleft ... \triangleleft Im(x_{n})$ where the $\triangleleft$ symbol means `is in the past light cone of'.
\end{enumerate}
We choose our integration contour in the form of a half disk with a diameter near the real $\sigma$ axis from $-R$ to $R$ and a semicircular region. (see Fig.~\ref{fig3}) Over this contour
\begin{equation}
\oint d\sigma (1-G(\sigma)) = 0
\end{equation}
Using the light cone OPE ~\eqref{eq:nullenergyinope} and our new coordinates, we have
\begin{equation}
G(\sigma) = 1 - \frac{\lambda_{T}}{N_{\delta}} \frac{\eta}{\sigma} \langle\overline{O(y=\delta)} \mathcal{E} O(y=\delta)\rangle
\end{equation}
where $N_{\delta} = \langle\overline{O(y=\delta)}O(y=\delta)\rangle$.
The sum of the integrals over the line and the semicircle is equal to 0. 
After doing the contour integral we are left with
\begin{equation}
i\langle\overline{O(y=\delta)} \mathcal{E} O(y=\delta)\rangle = \frac{N_{\delta}}{\pi\lambda_{T} \eta}\int_{-R}^{R} d\sigma Re(1-G(\sigma))
\end{equation}
Using property (1) of $G(\sigma)$, we have
\begin{equation}
i\langle\overline{O(y=\delta)} \mathcal{E} O(y=\delta)\rangle \geq 0
\end{equation}
Let $R$ be an operator which causes a rotation by $\frac{\pi}{2}$ in the $\tau-y$ plane where $\tau = \iota t$. In this process, the null contour along which $\mathcal{E}$ is computed is also rotated. This gives us
\begin{eqnarray}
(R.O)(t=-i\delta) &=& O(y=\delta) \cr
(R.O)^{\dagger}(t=i\delta) &=& \overline{O(y=\delta)} \cr
	\mathcal{E}' &=& \iota\mathcal{E}
	\end{eqnarray}
$\mathcal{E}'$ is the null energy computed along the rotated contour. Thus, we get
\begin{equation}
i\langle\overline{O(y=\delta)} \mathcal{E} O(y=\delta)\rangle = \langle(R.O)^{\dagger}(t=i\delta)\mathcal{E}(R.O)(t=-i\delta)\rangle
\end{equation}
Denoting $(R.O)$ by $\mathcal{O}$, we obtain
\begin{equation}
\langle\mathcal{O}^{\dagger}(t=i\delta)\mathcal{E}\mathcal{O}(t=-i\delta)\rangle \geq 0
\end{equation}
The above condition in a CFT is sufficient to say that $\mathcal{E}$ is a positive operator. This completes the proof of ANEC in a CFT.
%


\providecommand{\href}[2]{#2}%
\end{document}